\documentclass[a4paper]{PoS}
\usepackage{natbib/natbib}
\newcommand{\nodata}[1]{. . .}
\title{SOUSA's Swift Supernova Siblings}

\ShortTitle{Sibling Supernovae}

\author{\speaker{Peter J. Brown}\\
        George P. and Cynthia Woods Mitchell Institute for Fundamental Physics \& Astronomy, 
Texas A. \& M. University, Department of Physics and Astronomy, 
4242 TAMU, College Station, TX 77843, USA
        E-mail: \email{pbrown@physics.tamu.edu}}

\abstract{Swift has observed over three hundred supernovae in its first ten years.  
Photometry from the Ultra-Violet Optical Telescope (UVOT) is being compiled in the 
Swift Optical/Ultraviolet Supernovae Archive (SOUSA).  The diversity of supernovae leads to a wide dynamic range of intrinsic properties.  The intrinsic UV brightness of supernovae as a function of type and epoch allows one to understand the distance ranges at which Swift can reliably detect supernovae.  The large Swift sample also includes supernovae from the same galaxy as other Swift supernovae.  Through the first ten years, these families include 34 supernovae from 16 host galaxies (two galaxies have each hosted three Swift supernovae). 

}

\FullConference{Swift: 10 Years of Discovery,\\
		2-5 December 2014\\
		La Sapienza University, Rome, Italy }

\begin{document}

\section{Swift Supernova Observations}

In the first ten years of Swift operations \citep{Gehrels_etal_2004}, Swift has observed over three hundred supernovae.  All young supernovae (i.e. excluding supernova remnants and others which exploded before the launch of Swift) known by the author to have been observed by Swift are listed on the Swift Supernova website\footnote{http://swift.gsfc.nasa.gov/docs/swift/sne/swift\_sn.html.}.
The observed supernovae cover a wide variety of explosion and environment characteristics.    The UV data from the Ultra-Violet/Optical Telescope (UVOT; \citealp{Roming_etal_2005})  are being uniformly reduced and publicly released as the Swift Optical/Ultraviolet Supernova Archive (SOUSA).  The details of the photometric reduction are available in \citet{Brown_etal_2014_SOUSA}.  

As we look ahead to the next years of Swift supernova observations, results from the first ten years can guide the choice of future targets and observing strategies.  Most of the Swift supernova observations in the first ten years have focused on nearby supernovae with redshifts z$< $0.02.  Figure 1 shows the wide variety of absolute magnitudes in the mid-ultraviolet uvm2 filter.  Distance limits are estimated based on a limiting magnitude of 20 in the UVOT Vega system \citep{Poole_etal_2008,Breeveld_etal_2011}.  Fainter supernovae such as type Ib/c and subluminous type Ia supernovae can only be well observed within a redshift of z$\sim$0.005.  UV-bright type IIP and Ibn supernovae can be observed to z$\sim0.05$.  Some type Ia supernovae are also this bright, including so-called ``Super-Chandrasekhar'' type Ia supernovae \citep{Brown_etal_2014} and the peculiar SN~2011de \citep{Brown_2014}.  Very UV-bright type IIn can be observed to z$\sim0.2$ and the type II superluminous supernovae to z$\sim0.5$.  
Expanding the redshift range will increase the volume and numbers of some supernova types which can be observed with Swift.  In particular measuring the peak magnitudes of type Ia supernovae in the Hubble flow will reduce the uncertainties in the distances and absolute magnitudes, a problem with some of the nearby Swift supernova samples \citep{Brown_etal_2010}.

\begin{figure}
\includegraphics[width=1\textwidth]{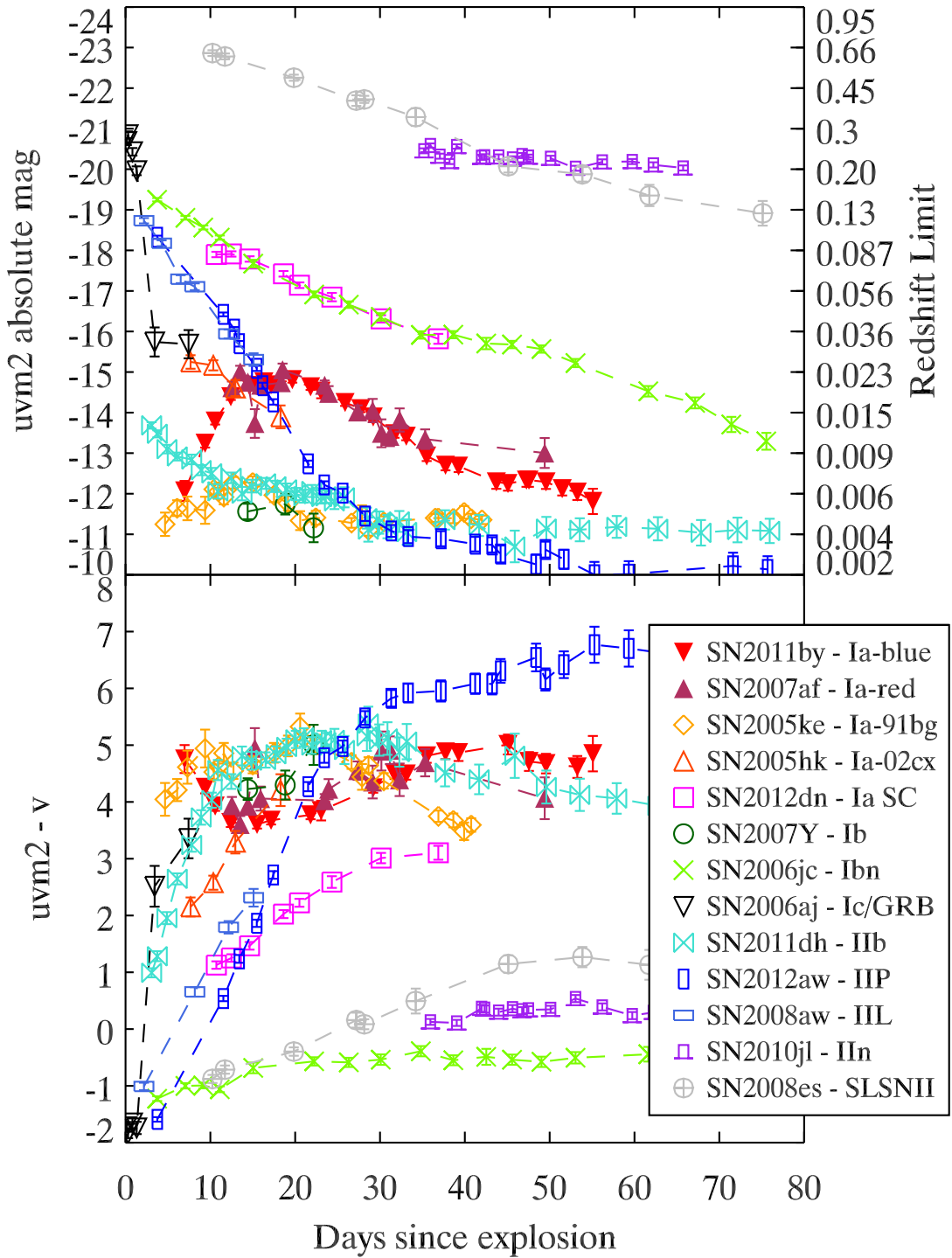}
\caption{ UV absolute magnitudes and UV-optical colors are plotted for prototypes of most supernova types to understand the detectability of supernovae based on the type and age and/or optical magnitude.  The redshift limits are based on a uvm2 limiting Vega magnitude of 20. }
\label{fig1}
\end{figure}

\section{Swift Supernova Siblings}

As the Swift supernova sample grows, so does the chance that some of the host galaxies have or will host another supernova.  We call these pairings of supernovae from the same hosts ``siblings.'' 
We reserve the term ``twins'' for supernovae which are nearly identical in many properties.  Supernova twins and siblings are more useful than just the sum of their individual utility.

Supernova twins are useful because their similar spectroscopic properties allow color differences to be traced to line of sight reddening, and their extinction corrected magnitudes can be used to infer distance differences \citep{Krisciunas_etal_2007,Krisciunas_etal_2009}.  The optical similarities of SNe 2011by and 2011fe led \citet{Foley_Kirshner_2013} to conclude that their UV differences were from metallicity.  \citet{Brown_etal_2015} find other effects such as density gradients also cause large UV differences while leaving the optical unscathed.  Since ``twin-ness'' can be very subjective, below we just focus on siblings, since host identification for nearby supernovae is a much more straightforward process.

\clearpage
Follow up observations of a supernova often result in a higher cadence of galaxy observations, so sometimes an additional supernova in the galaxy is discovered at a younger age than otherwise would have happened.  Observations of SN~1999ee in IC5179 led to the very early discovery of SN~1999ex and the first observations of the fading shock breakout of a Ib/c supernova \citep{Stritzinger_etal_2002}.  The best example is the discovery in Swift observations of SN~2007uy of the X-ray shock breakout of SN~2008D \citep{Soderberg_etal_2008}.  


Supernova siblings are important objects in the study of supernova rates as a function of supernova type and host galaxy morphology \citep{Anderson_Soto_2013}.  They are also important tests of the utility of global (rather than local) environment measurements and their correlation with rates or supernova properties.  
Sibling supernovae provide independent and joint constraints on the distance to the host galaxy \citep{Stritzinger_etal_2010} and a direct comparison of the absolute brightness of the siblings without the uncertainties on the distance.  In Table 1 we list the 16 host galaxies and the 34 supernovae they have hosted that were observed by Swift.  Other multiples have occurred during this period of time, but we are only listing those supernovae observed by Swift. 

\begin{table}
\caption{Swift Supernova Siblings}
\begin{tabular}{llllllll}
\hline\hline
{\bf Host Galaxy} & {\bf Redshift} & {\bf Supernova} & {\bf Type} & {\bf Supernova} & {\bf Type} & {\bf Supernova} & {\bf Type }\\ 
\hline
M51 & 0.002 & SN2005cs & IIP & SN2011dh & IIb & . . . & . . . \\
NGC1316     & 0.006 & SN2006dd & Ia & SN2006mr & Ia & . . . & . . . \\
NGC7364     & 0.016 & SN2006lc & Ib/c & SN2011im & Ia & . . . & . . . \\
MCG+05-43-16 & 0.017 & SN2007ck & IIP & SN2007co & Ia & . . . & . . . \\
NGC4039   & 0.005 & SN2007sr & Ia & SN2013dk & Ic	 & . . . & . . . \\		
NGC2770   & 0.006 & SN2007uy & Ib & SN2008D & Ib & . . . & . . . \\
ESO121-26 & 0.008 & SN2008M & IIP & SN2009mg & IIb & SN2012hr & Ia \\
NGC2765 & 0.013 & SN2008hv & Ia & ASASSN-13dd & Ia	 & . . . & . . . \\	
M61     & 0.005 & SN2008in & IIP & SN2014dt & Ia-02cx & . . . & . . . \\
NGC1954 & 0.010 & SN2010ko & Ia & SN2011fi & II  & SN2013ex & Ia \\
Arp299  & 0.010 & SN2010O & Ib & SN2010P & IIb & . . . & . . . \\
NGC1566 & 0.005 & SN2010el & Ia-02cx & ASASSN-14ha & II & . . . & . . . \\
NGC6240 & 0.024 & SN2010gp & Ia & PS1-14xw & Ia & . . . & . . . \\
NGC2207 & 0.009 & SN2010jp & IIn & SN2013ai & IIP & . . . & . . . \\
NGC5806 & 0.005 & SN2012P  & Ibc & iPTF13bvn & Ic & . . . & . . . \\
NGC7331 & 0.003 & SN2013bu & II & SN2014C & Ib & . . . & . . . \\[1ex]
\hline
\end{tabular}
\end{table}

\clearpage
\bibliographystyle{apj}
\bibliography{../bibtex}


\end{document}